\documentclass[12pt]{article}
\usepackage{epsf}
\title{ {\bf
Antisymmetric tensor unparticle and the radiative lepton flavor
violating decays}}
\author{\vspace{1cm}\\
        {\bf E. O. Iltan,}
        \thanks{E-mail address:
        eiltan@metu.edu.tr}
 \\
        Physics Department, Middle East Technical University \\
        Ankara, Turkey\\}

\date{}

\begin{document}
\setlength{\baselineskip}{24pt}
\maketitle
\setlength{\baselineskip}{7mm}
\begin{abstract}
We study the contribution of the tensor unparticle mediation to
the branching ratios of the radiative lepton flavor violating
decays and predict a restriction region for free parameters of the
scenario by using experimental upper limits. We observe that the
branching ratios of the radiative lepton flavor violating decays
are sensitive to the fundamental mass scales of the scenario and
to the scale dimension of antisymmetric tensor unparticle. We
obtain a more restricted set for the free parameters in the case
of the $\mu\rightarrow e \gamma$ decay.
\end{abstract}
\thispagestyle{empty}
\newpage
\setcounter{page}{1}
%
The radiative lepton flavor violating (LFV) decays $l_i\rightarrow
l_j\gamma$ reach great interest since their branching ratios (BRs)
in the framework of standard model (SM) are much below the
experimental upper limits, and, therefore, they are candidates to
search and to test more fundamental models beyond. The current
experimental upper limits of the BRs read BR $(\mu\rightarrow
e\gamma)=2.4\,(1.2)\times 10^{-12}\,(10^{-11})\,(\,\, 90\% CL)$
\cite{MEG} (\cite{Brooks}), BR $(\tau\rightarrow
e\gamma)=3.3\times 10^{-8}\,(\,\, 90\% CL)$ \cite{Aubert} and BR
$(\tau\rightarrow \mu\gamma)=4.4\times 10^{-8}\,(\,\, 90\% CL)$
\cite{Aubert}. There is an extensive theoretical work in the
literature in order to enhance BRs of these decays. They were
studied in the SM with the extended Higgs sector, so called two
Higgs doublet model (2HDM) \cite{Iltan1}-\cite{IltanLFVRS}, in
supersymmetric models \cite{Barbieri1}-\cite{Khalil}, in a model
independent way \cite{Chang}, in the framework of the 2HDM and the
supersymmetric model \cite{Paradisi}, in the SM including
effective operators coming from the possible unparticle effects
\cite{MuLinYan}-\cite{AndiHektor}, in little Higgs models
\cite{Illana}- \cite{Aguila}, in seesaw models \cite{GangHe}
-\cite{Deppisch}, in models with A(4) and S(4) flavor symmetries
\cite{Ding}, using the effective field theory  with Higgs
mediation \cite{Aranda}, in the Higgs triplet model
\cite{Akeroyd}, in the framework of Higgs-induced lepton flavor
violation \cite{Goudelis}.

In the present work, we consider the contribution of the
antisymmetric tensor unparticle mediation to the BRs of the
radiative LFV decays (see \cite{Tae} for the contribution of the
antisymmetric tensor unparticle mediation to the muon anomalous
magnetic dipole moment, to the electroweak precision observable
$S$, its effects in $Z$ invisible decays and  see
\cite{IltanTensUnp} for the contribution of the antisymmetric
tensor unparticle mediation to the lepton electric dipole moment).
Unparticles \cite{Georgi1, Georgi2}, being massless due to the
scale invariance and having non integral scaling dimension $d_U$
around the scale $\Lambda_U\sim 1.0\,TeV$, come out with the
interaction of SM-ultraviolet sector at some scale $M_U$:
\begin{equation}
{\cal{L}}_{eff}= \frac{C_n}{M_U^{d_{UV}+n-4}}\,O_{SM}\,O_{UV}\,,
\label{eff1}
\end{equation}
where $d_{UV}$ is the scaling dimension of the UV operator
\cite{BankZaks}. Around the scale $\Lambda_U$ the effective
interaction becomes \cite{Rajaraman}
\begin{equation}
{\cal{L}}_{eff}=
\frac{C^i_n}{\Lambda_n^{d_{U}+n-4}}\,O_{SM,i}\,O_{U}\,.
\label{eff2}
\end{equation}
Here $O_{SM,i}$ is type $i$ SM operator, $n$ is its scaling
dimension and $\Lambda_n$ is the mass scale (see \cite{Tae,
Rajaraman} for details) which reads
\begin{equation}
\Lambda_n=\Bigg(\frac{M_U^{d_{UV}+n-4}}{\Lambda_U^{d_{UV}-d_U}}\Bigg)^{
\frac{1}{d_U+n-4}} \,. \label{eff3}
\end{equation}

The antisymmetric tensor unparticle mediation induces these LFV
decays at tree level and we consider the case that the scale
invariance is broken at some scale $\mu$ after the electroweak
symmetry breaking (see for example \cite{PJFox, Kikuchi} for a
possible interaction which causes that scale invariance is
broken). The effective lagrangian \cite{Tae} which can drive the
radiative LFV decays is
\begin{eqnarray}
{\cal{L}}_{eff}&=&\frac{g'\,\lambda_B}{\Lambda_2^{d_U-2}}\,B_{\mu\nu}\,
O^{\mu\nu}_U+\frac{g\,\lambda_W}{\Lambda_4^{d_U}}\,(H^\dagger\,\tau_a
\,H)\, W^a_{\mu\nu}\, O^{\mu\nu}_U  \nonumber \\
&+& \frac{\lambda_{ij}}{\Lambda_4^{d_U}}\,y_{ij}\, \bar{l}_i\,H
\,\sigma_{\mu\nu}\,l_j\, O^{\mu\nu}_U\, , \label{lagrangiantensor}
\end{eqnarray}
where $l_{i(j)}$ is the lepton field,  $H$ is the Higgs doublet,
$g$ and $g'$ are weak couplings, $\lambda_B$, $\lambda_W$ and
$\lambda_{ij}$ are the unparticle-field tensor and
unparticle-lepton-lepton couplings, $B_{\mu\nu}$ is the field
strength tensor of the $U(1)_Y$ gauge boson with
$B_\mu=c_W\,A_\mu+s_W\,Z_\mu$ and $W^a_{\mu\nu}$, $a=1,2,3$, are
field strength tensors of the $SU(2)_L$ gauge bosons with
$W^3_\mu=s_W\,A_\mu-c_W\,Z_\mu$ and $A_\mu$ ($Z_\mu$) is photon (Z
boson) field. The couplings $y_{ij}$ are responsible for the LF
violation and after the electroweak symmetry breaking we introduce
modified couplings $\xi_{ij}=\frac{v}{\sqrt{2}}\,y_{ij}$ which
respect the mass hierarchy of charged leptons. The process
$l_i\rightarrow l_j\gamma$ appears in the tree level with the
communication of two vertices\footnote{The first vertex arises
from the last term of the effective lagrangian and leads to the
$l_i\rightarrow l_j$ transition. The second one arises from the
first and second terms of the effective lagrangian and results in
the $O^{\mu\nu}_U \rightarrow A_\nu$ transition }
$\lambda_{ij}\,\frac{\xi_{ij}}{\Lambda_4^{d_U}}\,
\bar{l}_i\,\sigma_{\mu\nu}\,l_j\,O^{\mu\nu}_U $ and $\Big(
\,2\,i\,\frac{g'\,c_W\,\lambda_B}{\Lambda_2^{d_U-2}}-
i\,\frac{g\,v^2\,s_W\,\lambda_W}{2\,\Lambda_4^{d_U}}\, \Big)
\,k_\mu\,\epsilon_\nu \,O^{\mu\nu}_U$,
by the antisymmetric tensor unparticle propagator (see Appendix
and eq.(\ref{propagatormu})) and the matrix element of this
process reads
\begin{eqnarray}
M=a_{ij}\,\bar{l}_i\,\sigma_{\mu\nu}\,l_j\,k_\mu\,\epsilon_\nu \,
, \label{MatrixEl}
\end{eqnarray}
where
\begin{eqnarray}
a_{ij}=\frac{i\,e\,\mu^{2\,(d_U-2)}\,\,A_{d_U}\,\lambda_{ij}\,\xi_{ij}}
{sin\,(d_U\pi)\,\Lambda_4^{d_U}}\,\Bigg(\frac{\lambda_B}{\Lambda_2^{d_U-2}}-
\frac{v^2\,\lambda_W}{4\,\Lambda_4^{d_U}} \Bigg)\, . \label{aij}
\end{eqnarray}
Finally the decay width $\Gamma(l_i\rightarrow l_j\gamma)$ becomes
\begin{eqnarray}
\Gamma(l_i\rightarrow
l_j\gamma)=\frac{1}{8\,\pi}\,m_i^3\,|a_{ij}|^2 \, , \label{DW}
\end{eqnarray}
where $m_i$ is the mass of incoming lepton. Notice that, in this
expression, we ignore the mass of outgoing one.
\newpage
%
%
{\Large \textbf{Discussion}}
\\

\begin{figure}[htb]
\vskip 6.0truein \centering \epsfxsize=5.8in
\leavevmode\epsffile{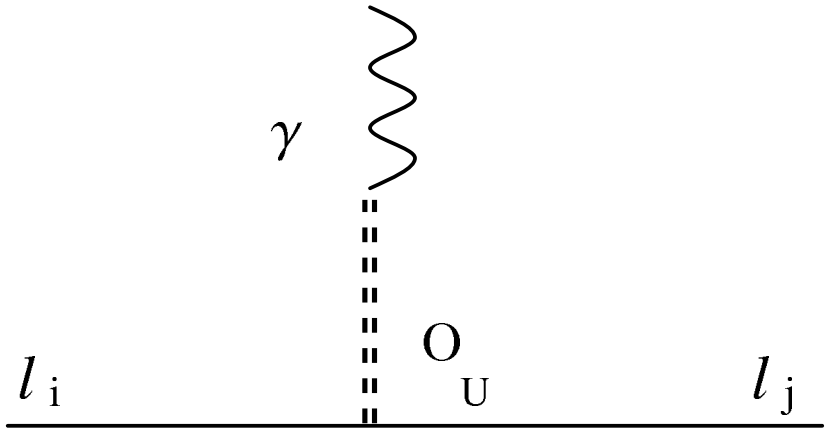} \vskip -16.0truein
\caption[]{Tree level diagram contributing to the $l_i\rightarrow
l_j\, \gamma$ decay due to antisymmetric tensor unparticle. Wavy
(solid) line represents the electromagnetic field (lepton field)
and double dashed line the antisymmetric tensor unparticle field.}
\label{fig1}
\end{figure}

In this section we study the intermediate antisymmetric tensor
unparticle contribution to the radiative LFV decays
$l_i\rightarrow l_j\gamma$ which exist at tree level (see
Fig.\ref{fig1}). There are various free parameters in this
scenario and we restrict them by using the current experimental
upper limits of BRs of LFV decays. Now, we would like to present
the free parameters and discuss the restrictions predicted. The SM
sector interacts with the UV one and it appears as unparticle
sector at a lower scale. The corresponding UV (unparticle)
operator $O_{UV}$ ($O_{U}$) has the scaling dimension $d_{UV}$
($d_{U}$) which is among the free parameters. We choose the scale
dimension $d_U$ in the range $1< d_U <2$. Notice that the scale
dimension must satisfy $d_U>2$ for antisymmetric tensor unparticle
in order not to violate the unitarity \cite{Grinstein}. Our
assumption is based on the fact that the scale invariance is
broken at some scale $\mu$ and one reaches to the particle sector.
This results in a relaxation  on the values of $d_U$ and we choose
$d_U$ in the range $1< d_U <2$ so that the propagator for particle
sector is obtained when $d_U$ tends to one. Furthermore we choose
the numerical value of $d_{UV}$ as $d_{UV}=3$ which satisfies
$d_{UV}>d_U$ (see \cite{Rajaraman}). The SM-ultraviolet sector
interaction scale $M_U$, the SM-unparticle sector interaction
scale $\Lambda_U$ and the scale $\mu$ which is the one that scale
invariance is broken belong to the free parameter set of the
present scenario. Here we choose $\mu\sim 1.0\,GeV$ and predict
the restrictions for the others by using the experimental upper
limits of LFV decays. Finally, for the couplings $\lambda_B$,
$\lambda_W$ and $\lambda_{ij}$ we consider
$\lambda_B=\lambda_W=\lambda_{ij}=1$ and for $\xi_{ij}$ we respect
the mass hierarchy of charged leptons, namely we choose
$\xi_{\tau\mu}>\xi_{\tau e}>\xi_{\mu e}$ and we take
$\xi_{\tau\mu}=0.1\,GeV$, $\xi_{\tau e}=0.01\,GeV$ and $\xi_{\mu
e}=0.001\,GeV$ in our numerical calculations.

\begin{figure}[htb]
\vskip -3.0truein \centering \epsfxsize=6.8in
\leavevmode\epsffile{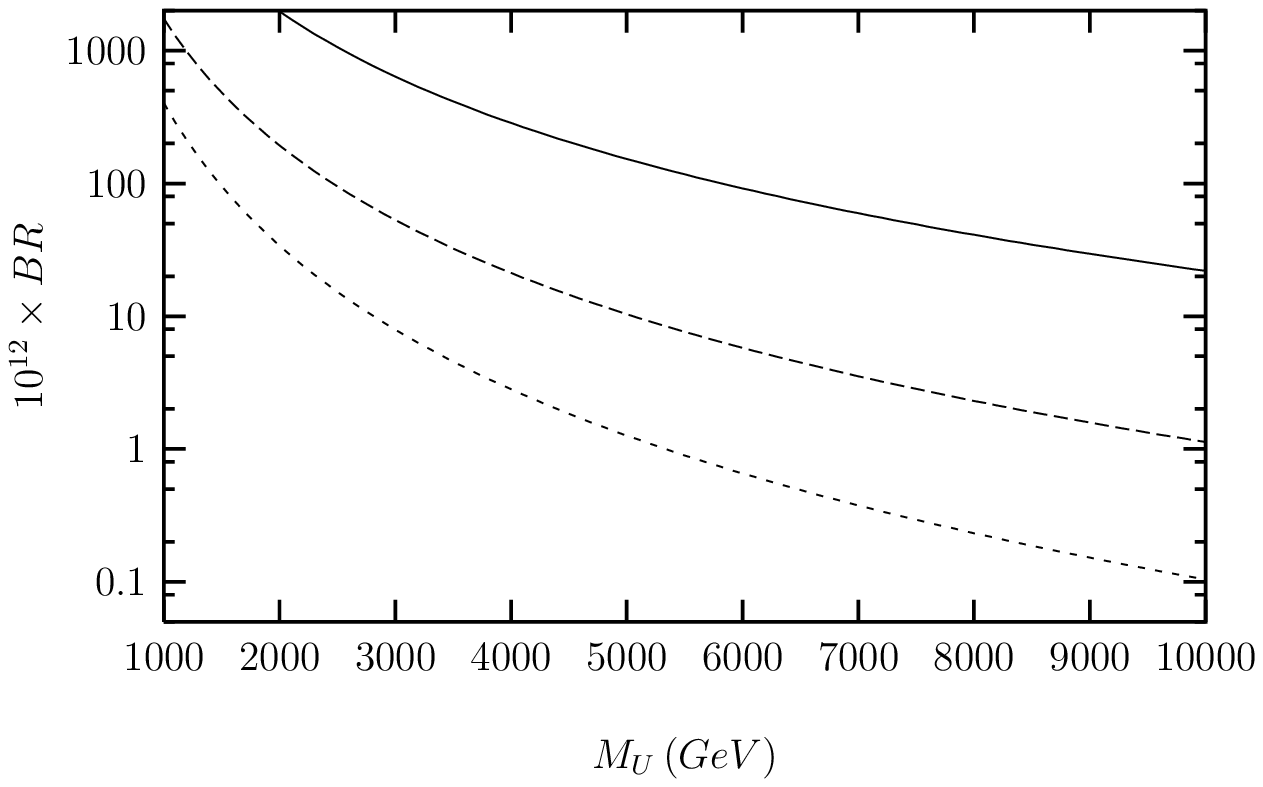} \vskip -3.0truein
\caption[]{$M_U$ dependence of the BR $(\mu\rightarrow e\,
\gamma)$ for $r_U=0.1$. Here, the solid (long dashed-short dashed)
line represents the BR for $d_U=1.7\,(1.8-1.9)$.}
\label{muegrt0001Mmu}
\end{figure}

\begin{figure}[htb]
\vskip -3.0truein \centering \epsfxsize=6.8in
\leavevmode\epsffile{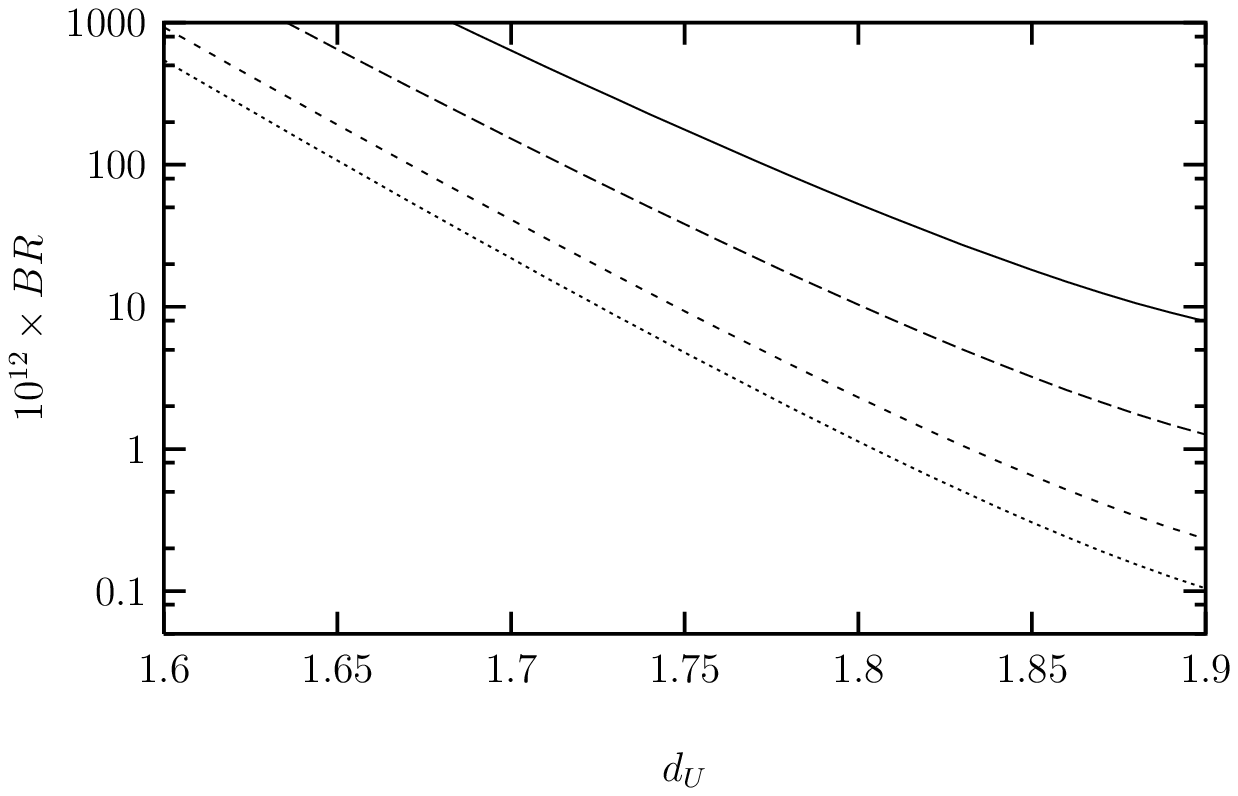} \vskip -3.0truein
\caption[]{$d_U$ dependence of the BR $(\mu\rightarrow e\,
\gamma)$ for $r_U=0.1$. Here the solid (long dashed-short
dashed-dotted) line represents the BR for $M_U=3000\,
(5000-8000-10000)\, GeV$.} \label{muegrt0001du}
\end{figure}

In Fig.\ref{muegrt0001Mmu}, we present the BR$(\mu\rightarrow e\,
\gamma$) with respect to the mass scale $M_U$ for
$r_U=\frac{\Lambda_U}{M_U}=0.1$. Here, the solid (long
dashed-short dashed) line represents the BR for
$d_U=1.7\,(1.8-1.9)$. We observe that the BR is sensitive to the
mass scale $M_U$ especially for the large values of the scale
dimension $d_U$ and it decreases almost three orders in the range
of $2000\, GeV < M_U < 10000\, GeV$. The experimental upper limit
is reached for $d_U\sim 1.8\, (1.9)$ and $M_U \sim 8000 \,
(4500)\, GeV$. Fig.\ref{muegrt0001du} is devoted to the
BR$(\mu\rightarrow e\, \gamma$) with respect to the scale
parameter $d_U$ for $r_U=0.1$. Here the solid (long dashed-short
dashed-dotted) line represents the BR for $M_U=3000\,
(5000-8000-10000)\, GeV$. The BR strongly depends on $d_U$ and
decreases with the increasing values of $d_U$. The experimental
upper limit is observed in the range of $d_U\sim 1.78-1.88$ for
$M_U\sim 5000-10000\, GeV$.

\begin{figure}[htb]
\vskip -3.0truein \centering \epsfxsize=6.8in
\leavevmode\epsffile{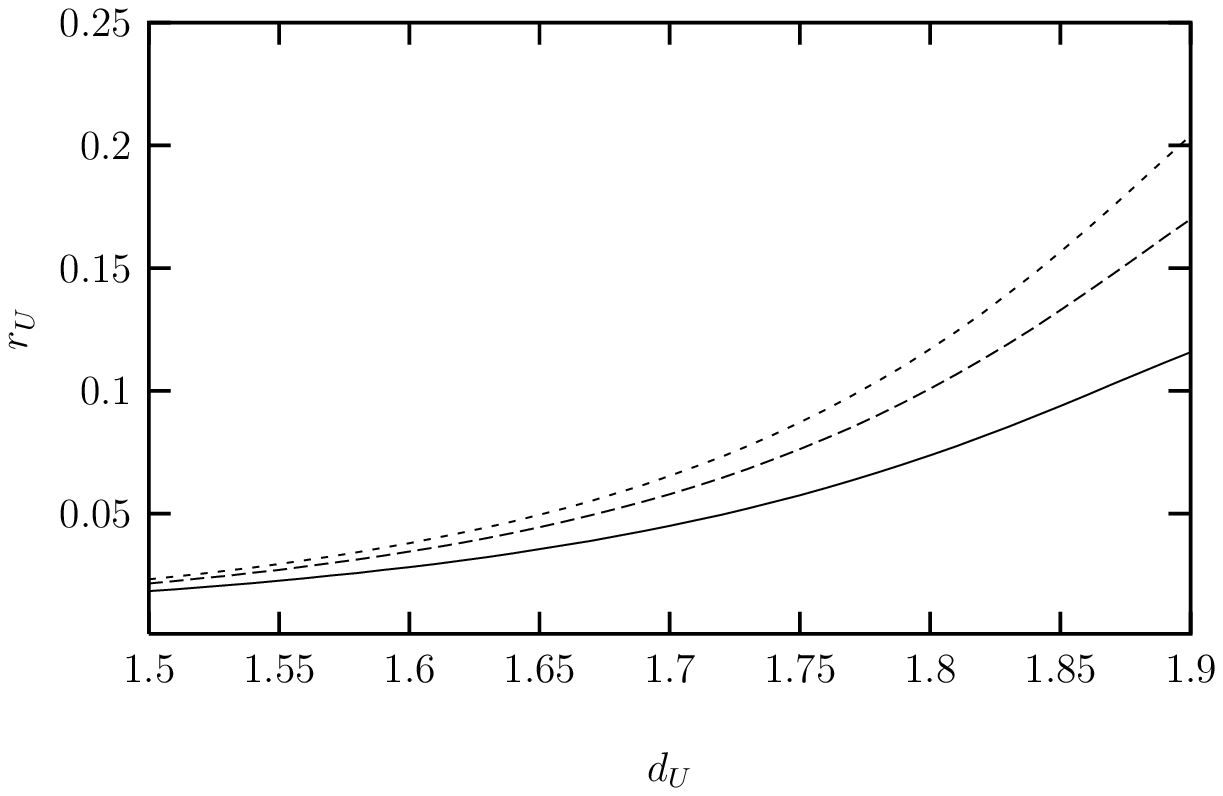} \vskip -3.0truein
\caption[]{$r_U$ with respect to $d_U$ for BR$(\mu\rightarrow e\,
\gamma)=2.4\times 10^{-12}$. Here the solid (long dashed-short
dashed) line represents $r_U$ for $M_U=5000\,(8000-10000)\,GeV$}
\label{muegrtdu}
\end{figure}

In Fig.\ref{muegrtdu} we show the parameter $r_U$ with respect to
$d_U$ for BR$(\mu\rightarrow e\, \gamma)=2.4\times 10^{-12}$. Here
the solid (long dashed-short dashed) line represents $r_U$ for
$M_U=5000\, (8000-10000)\,GeV$. We see that the scale dimension
$d_U$ and $r_U$ can take values in the range $1.73-1.90$ and
$0.05-0.12$, respectively for $M_U=5000\,GeV$. For
$M_U=10000\,GeV$ we have the range $1.65-1.9$ for $d_U$ and
$0.05-0.20$ for $r_U$.

\begin{figure}[htb]
\vskip -3.0truein \centering \epsfxsize=6.8in
\leavevmode\epsffile{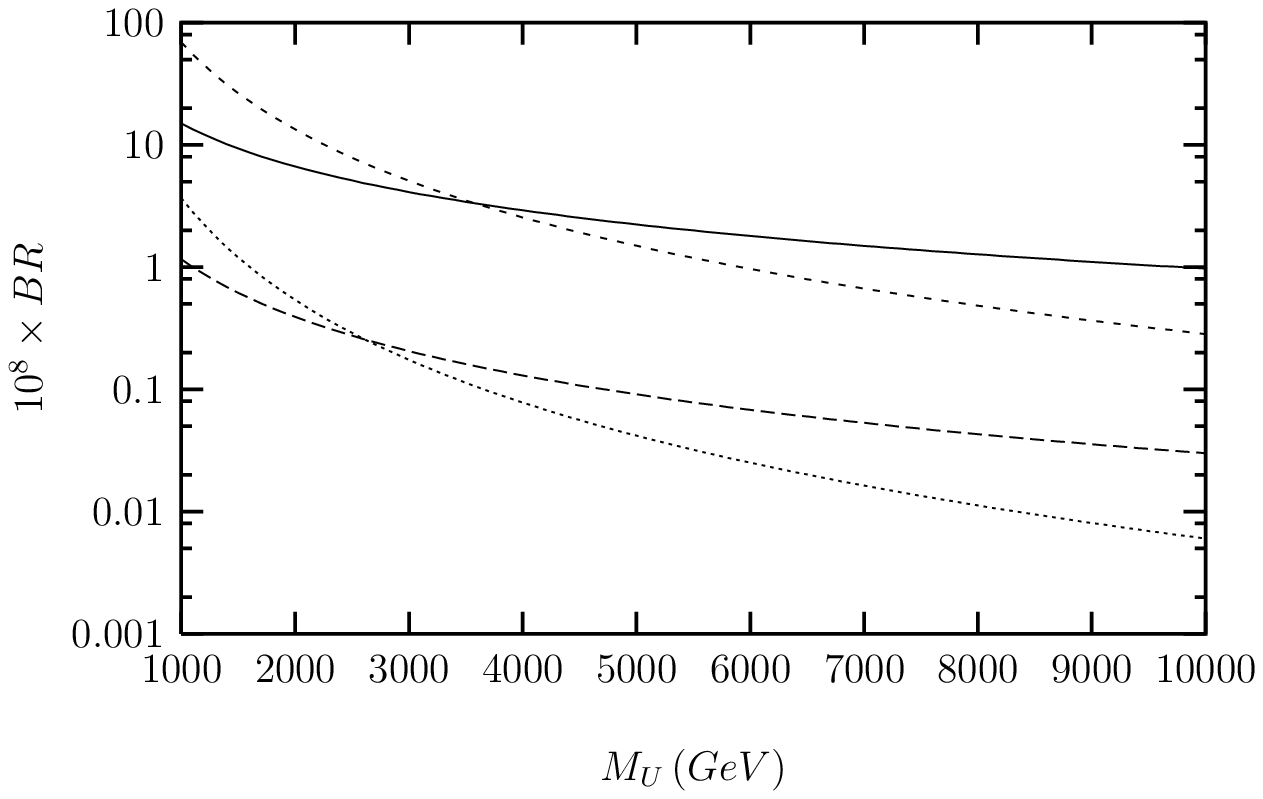} \vskip -3.0truein
\caption[]{$M_U$ dependence of the BR $(\tau\rightarrow e\,
\gamma)$. Here, the solid (long dashed-short dashed-dotted) line
represents the BR for $r_U=0.1$, $d_U=1.3$ ($r_U=0.1$,
$d_U=1.4$-$r_U=0.5$, $d_U=1.6$-$r_U=0.5$, $d_U=1.7$).}
\label{mtauegrt01and05Mmu}
\end{figure}

Fig.\ref{mtauegrt01and05Mmu} represents the BR$(\tau\rightarrow
e\, \gamma$) with respect to the mass scale $M_U$. Here, the solid
(long dashed-short dashed-dotted) line represents the BR for
$r_U=0.1$, $d_U=1.3$ ($r_U=0.1$, $d_U=1.4$-$r_U=0.5$,
$d_U=1.6$-$r_U=0.5$, $d_U=1.7$). It is observed that the
sensitivity of the BR to the mass scale $M_U$ increases with the
increasing values of the ratio $r_U$. The experimental upper limit
is reached for $r_U=0.1$ and $d_U\sim 1.3$ when the mass scale
$M_U$ reads $M_U \sim 4000 \, GeV$. For $r_U=0.5$ one reaches the
experimental limit in the case of $d_U\sim 1.6$ and $M_U \sim 4000
\, GeV$. Fig.\ref{mtauegrt01and05Mmu} shows the
BR$(\tau\rightarrow e\, \gamma$) with respect to the scale
parameter $d_U$ for $r_U=0.1$. Here the solid (long dashed-short
dashed) line represents the BR for $M_U=2000\, (5000-10000)\,GeV$.
The BR strongly depends on $d_U$ and decreases with the increasing
values of $d_U$ similar to the $\mu\rightarrow e\, \gamma$ decay.
One reaches the experimental upper limit in the range of $d_U\sim
1.26-1.32$ for $M_U\sim 2000-10000\,GeV$.

Fig.\ref{mtauegrtdu} is devoted to the parameter $r_U$ with
respect to $d_U$ for BR$(\tau\rightarrow e\, \gamma)=3.3\times
10^{-8}$. Here the solid (long dashed-short dashed) line
represents $r_U$ for $M_U=5000\, (8000-10000)\,GeV$. This figure
shows that the experimental upper limit is reached for
$M_U=5000\,GeV$ if the scale dimension $d_U$ and $r_U$ can take
values in the range $1.10-1.68$ and $0.05-1.00$, respectively. For
$M_U=10000\,GeV$ we have the range $1.10-1.64$ for $d_U$ and
$0.05-1.00$ for $r_U$.

In  Fig.\ref{taumugrt01Mmu}, we present the BR$(\tau\rightarrow
\mu\, \gamma$) with respect to the mass scale $M_U$ for $r_U=0.1$.
Here, the solid (long dashed-short dashed-dotted) line represents
the BR for $d_U=1.4\,(1.5-1.6-1.7)$. We observe that the BR
decreases more than one order in the range of $2000\, GeV < M_U <
10000\, GeV$. The experimental upper limit is reached for $d_U\sim
1.4$ and $M_U \sim 9000\, GeV$. Fig.\ref{taumugrt01du} represents
the BR$(\tau\rightarrow \mu\, \gamma)$ with respect to the scale
parameter $d_U$ for $r_U=0.1$. Here the solid (long dashed-short
dashed) line represents the BR for $M_U=2000\, (5000-10000)\,GeV$.
The experimental upper limit of the BR is observed in the range of
$d_U\sim 1.38-1.47$ for $M_U\sim 2000-10000\, GeV$.

In Fig. \ref{taumugrtdu} we show the parameter $r_U$ with respect
to $d_U$ for BR$(\tau\rightarrow \mu\, \gamma)=4.4\times 10^{-8}$.
Here the solid (long dashed-short dashed) line represents $r_U$
for $M_U=2000\, (5000-10000)\,GeV$. We see that the scale
dimension $d_U$ and $r_U$ can take values in the range
$1.48-1.80\,(1.3-1.8)$ and $0.10-0.55\,(0.05-1.00)$, respectively
for $M_U=2000\, (5000)\,GeV$. For $M_U=10000\,GeV$ we have the
range $1.30-1.75$ for $d_U$ and $0.05-1.00$ for $r_U$.

As a summary, the BRs of radiative LFV decays are sensitive to the
mass scale $M_U$ especially for the large values of the scale
dimension $d_U$ and this sensitivity increases with the increasing
values of the ratio $r_U$. The experimental upper limit of the
BR$(\mu\rightarrow e\, \gamma$) can be reached for $r_U\sim 0.1$,
$d_U > 1.7$ and for larger values of $M_U$, namely $M_U \sim
7000-9000 \, GeV$. For BR$(\tau\rightarrow e\, \gamma$) one
reaches the experimental upper limit for $r_U\sim 0.1$, $d_U\sim
1.3$ and $M_U > 2000\,GeV$ . If we consider the $\tau\rightarrow
\mu\, \gamma$ decay the experimental upper limit of BR is obtained
for $r_U\sim 0.1$, when $d_U$ is in the range $d_U \sim 1.4-1.5$
and $M_U> 2000\,GeV$. We see that the free parameters of this
scenario are more restricted if the $\mu\rightarrow e\, \gamma$
decay is considered. However the future more accurate measurements
of the upper limits of the LFV decays make it possible to obtain a
more restricted range for the free parameters of this scenario and
they stimulate to search the role and the nature of unparticle
physics which is a candidate to drive the lepton flavor violation.
\newpage
{\Large \textbf{Appendix}}
\\ \\
The scalar unparticle propagator reads \cite{Georgi2, Cheung2}
\begin{eqnarray}
\int\,d^4x\,
e^{ip.x}<0|T\Big(O_U(x)\,O_U(0)\Big)0>=i\frac{A_{d_U}}{2\,\pi}\,
\int_0^{\infty}\!\!\!
ds\,\frac{s^{d_U-2}}{p^2-s+i\epsilon}\!=i\,\frac{A_{d_U}}
{2\,sin\,(d_U\pi)}\,(-p^2-i\epsilon)^{d_U-2} \label{propagators}
\end{eqnarray}
where the factor $A_{d_U}$ is
\begin{eqnarray}
A_{d_U}=\frac{16\,\pi^{5/2}}{(2\,\pi)^{2\,d_U}}\,
\frac{\Gamma(d_U+\frac{1}{2})} {\Gamma(d_U-1)\,\Gamma(2\,d_U)} \,
. \label{Adu}
\end{eqnarray}
Now the tensor unparticle propagator is obtained by using the
projection operator $\Pi^{\mu\nu\alpha\beta}$
\begin{eqnarray}
\Pi_{\mu\nu\alpha\beta}=
\frac{1}{2}(g_{\mu\alpha}\,g_{\nu\beta}-g_{\nu\alpha}\,g_{\mu\beta})
\, , \label{projection}
\end{eqnarray}
with the transverse and the longitudinal parts
\begin{eqnarray}
\Pi^T_{\mu\nu\alpha\beta}=
\frac{1}{2}(P^T_{\mu\alpha}\,P^T_{\nu\beta}-P^T_{\nu\alpha}\,P^T_{\mu\beta})\,
, \,\,\,\,\,\,
\Pi^L_{\mu\nu\alpha\beta}=\Pi_{\mu\nu\alpha\beta}-\Pi^T_{\mu\nu\alpha\beta}
\, , \label{projectionTL}
\end{eqnarray}
where $P^T_{\mu\nu}=g_{\mu\nu}-p_{\mu}\,p_{\nu}/{p^2}$ (see for
example \cite{Tae} and references therein) and the propagator of
antisymmetric tensor unparticle becomes
\begin{eqnarray}
\int\,d^4x\,
e^{ipx}\,<0|T\Big(O^{\mu\nu}_U(x)\,O^{\alpha\beta}_U(0)\Big)0>=
i\,\frac{A_{d_U}}
{2\,sin\,(d_U\pi)}\,\Pi^{\mu\nu\alpha\beta}(-p^2-i\epsilon)^{d_U-2}
\, . \nonumber
\end{eqnarray}
On the other hand the propagator is modified if the scale
invariance broken at a certain scale and this modification is
model dependent. Following the  the simple model \cite{PJFox,
ARajaraman, ADelgado} we take
\begin{eqnarray}
\int\,d^4x\,
e^{ipx}\,<0|T\Big(O^{\mu\nu}_U(x)\,O^{\alpha\beta}_U(0)\Big)0>=
i\,\frac{A_{d_U}}
{2\,sin\,(d_U\pi)}\,\Pi^{\mu\nu\alpha\beta}(-(p^2-\mu^2)-i\epsilon)^{d_U-2}
\, . \label{propagatormu}
\end{eqnarray}
where $\mu$ is the scale that the scale invariance broken and the
particle sector comes out.
\newpage
\newpage
\newpage
\begin{figure}[htb]
\vskip -3.0truein \centering \epsfxsize=6.8in
\leavevmode\epsffile{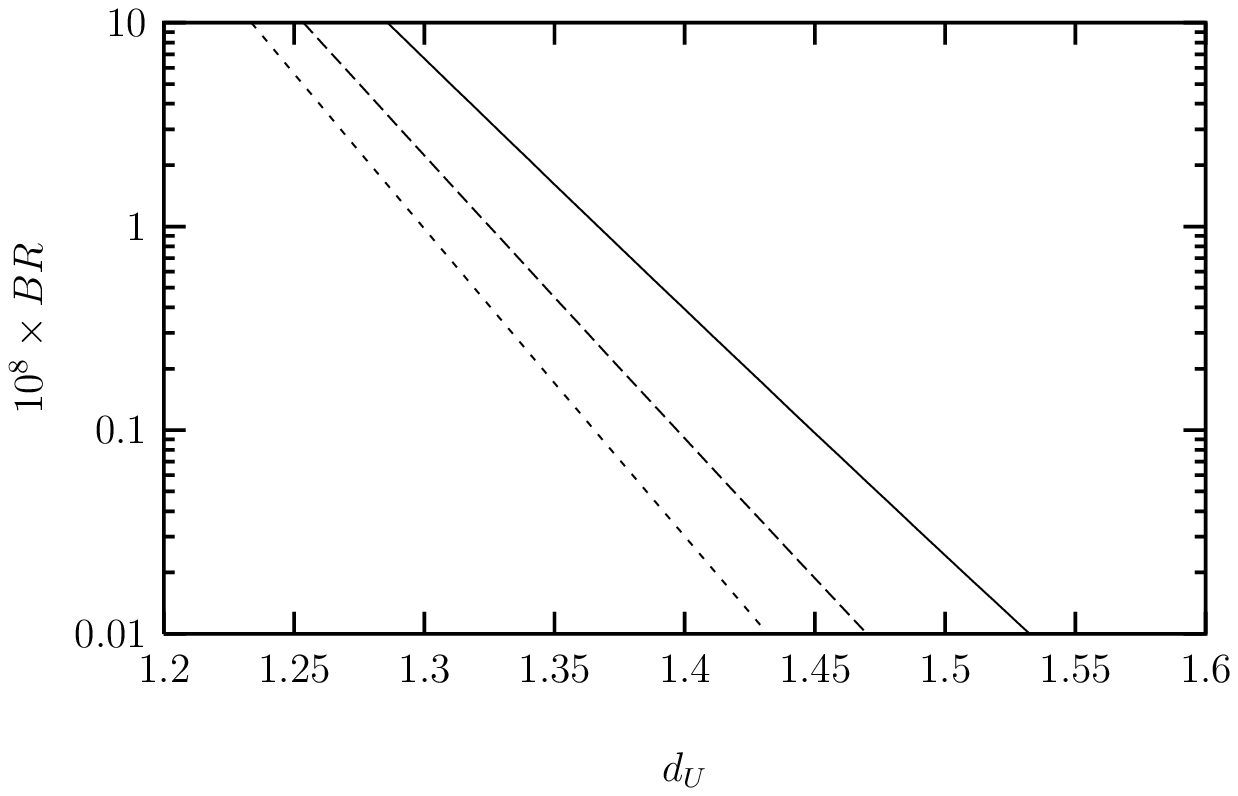} \vskip -3.0truein
\caption[]{$d_U$ dependence of the BR $(\tau\rightarrow e\,
\gamma)$  for $r_U=0.1$. Here the solid (long dashed-short dashed)
line represents the BR for $M_U=2000\, (5000-10000)\,GeV$.}
\label{mtauegrt01du}
\end{figure}
\begin{figure}[htb]
\vskip -3.0truein \centering \epsfxsize=6.8in
\leavevmode\epsffile{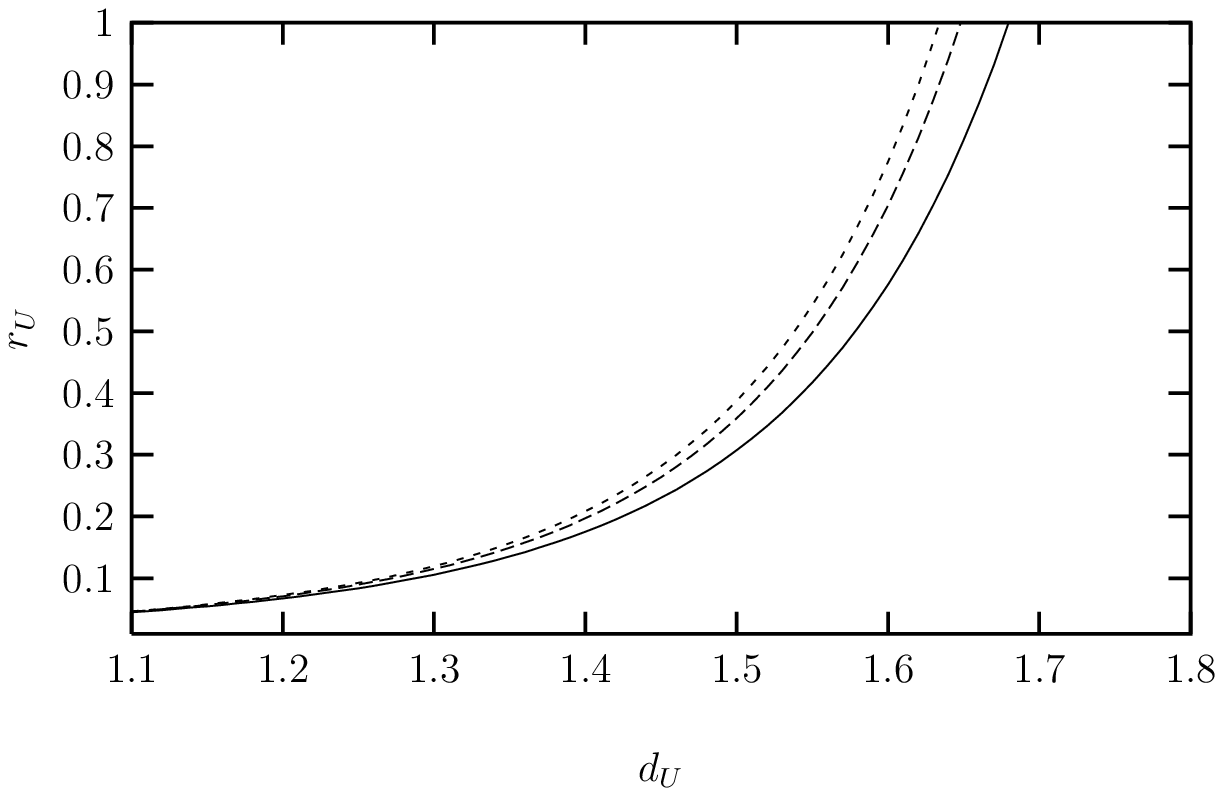} \vskip -3.0truein
\caption[]{$r_U$ with respect to $d_U$ for BR$(\tau\rightarrow e\,
\gamma)=3.3\times 10^{-8}$. Here the solid (long dashed-short
dashed) line represents $r_U$ for $M_U=5000\, (8000-10000)\,GeV$.
} \label{mtauegrtdu}
\end{figure}
\begin{figure}[htb]
\vskip -3.0truein \centering \epsfxsize=6.8in
\leavevmode\epsffile{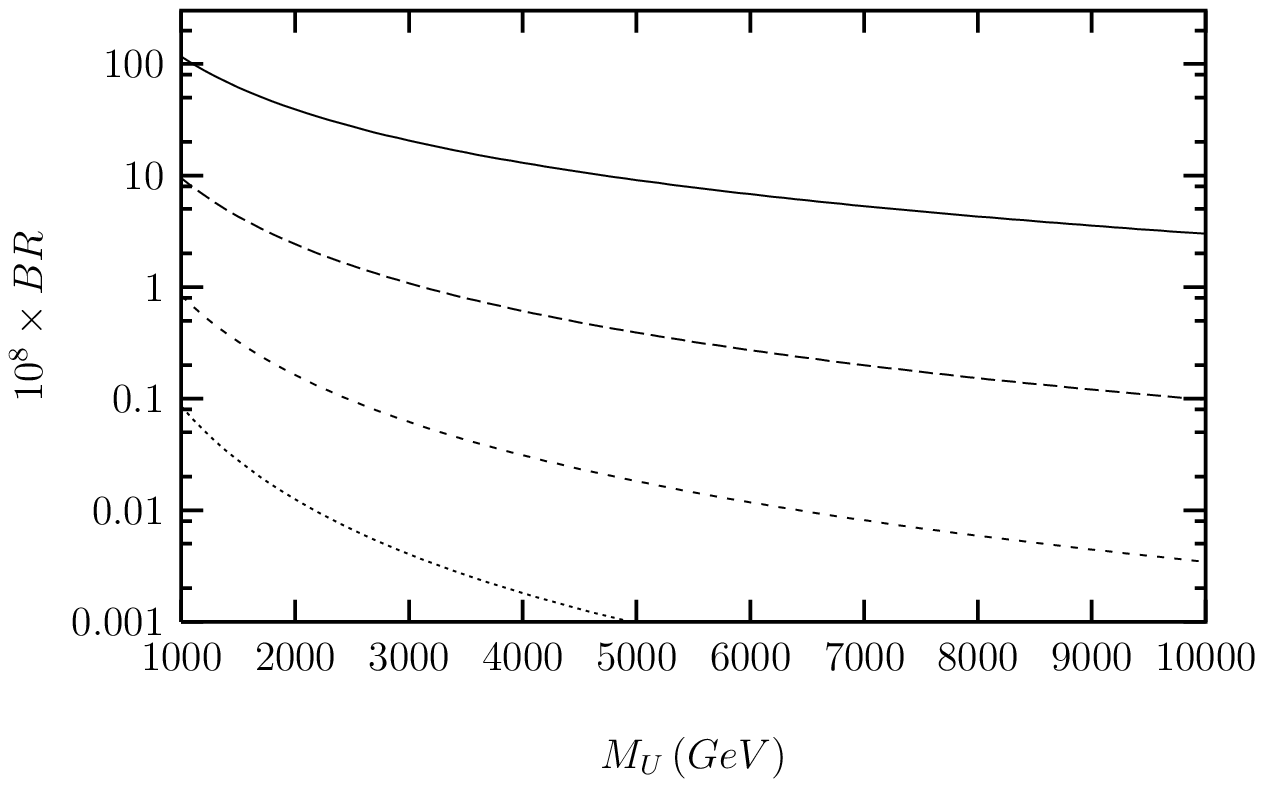} \vskip -3.0truein
\caption[]{ $M_U$ dependence of the BR$(\tau\rightarrow \mu\,
\gamma)$. Here, the solid (long dashed-short dashed-dotted) line
represents the BR for $d_U=1.4\,(1.5-1.6-1.7)$. }
\label{taumugrt01Mmu}
\end{figure}
\begin{figure}[htb]
\vskip -3.0truein \centering \epsfxsize=6.8in
\leavevmode\epsffile{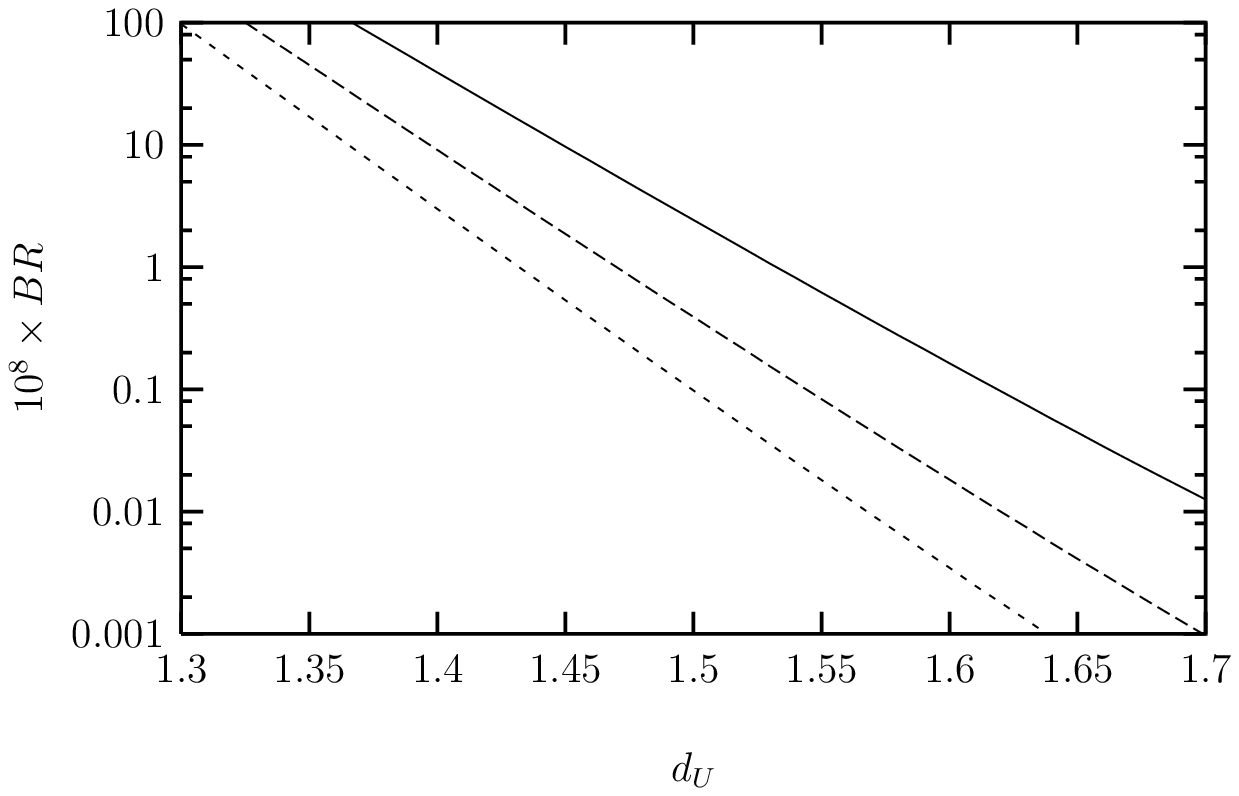} \vskip -3.0truein
\caption[]{$d_U$ dependence of the BR$(\tau\rightarrow \mu\,
\gamma)$ for $r_U=0.1$. Here the solid (long dashed-short dashed)
line represents the BR for $M_U=2000\, (5000-10000)\,GeV$.}
\label{taumugrt01du}
\end{figure}
\begin{figure}[htb]
\vskip -3.0truein \centering \epsfxsize=6.8in
\leavevmode\epsffile{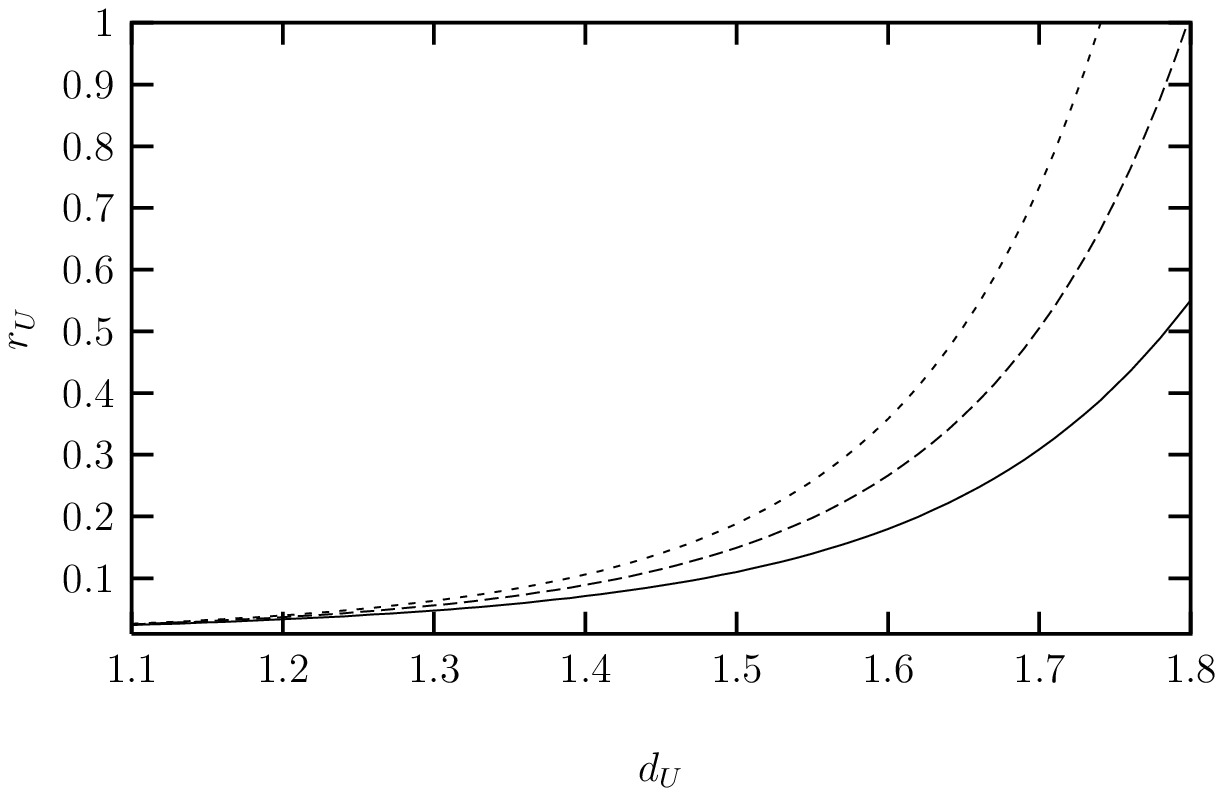} \vskip -3.0truein
\caption[]{$r_U$ with respect to $d_U$ for BR$(\tau\rightarrow
\mu\, \gamma)=4.4\times 10^{-8}$. Here the solid (long
dashed-short dashed) line represents $r_U$ for $M_U=2000\,
(5000-10000)\,GeV$. } \label{taumugrtdu}
\end{figure}

\begin{thebibliography}{1}
%
\bibitem{MEG} J. Adam et.al., MEG Collaboration, hep-ex/1107.5547, (2011).
%
\bibitem{Brooks} M. L. Brooks et. al., MEGA Collaboration,
{\it Phys. Rev. Lett.} {\bf 83}, 1521 (1999).
%
\bibitem{Aubert} B. Aubert et. al., BABAR Collaboration,
BABAR-PUB-09/026, SLAC-PUB-13753, {\it Phys. Rev. Lett.} {\bf 104}
021802, (2010).
%
\bibitem{Iltan1} E. O. Iltan, {\it Phys. Rev.} {\bf D64} 115005, (2001).
%
\bibitem{Iltan11} E. O. Iltan,{\it Phys. Rev.} {\bf D64} 013013, (2001)
%
\bibitem{Diaz} R. Diaz, R. Martinez and J-Alexis Rodriguez,
Phys.Rev. D63 095007, (2001).
%
\bibitem{IltanExtrDim} E. O. Iltan, {\it JHEP} {\bf 0402} 20, (2004).
%
\bibitem{IltanExtrDim1} E. O. Iltan, {\it JHEP} {\bf 0408} 20, (2004)
%
\bibitem{IltanLFVSplit} E. O. Iltan, {\it Mod. Phys. Lett.} {\bf A22} 819,
(2007).
%
\bibitem{IltanLFVRS} E. O. Iltan, {\it Int. J. Mod. Phys.} {\bf
A23} 1055, (2008).
%
\bibitem{Barbieri1} R. Barbieri and L. J. Hall,
{\it Phys. Lett.} {\bf B338} 212, (1994).
%
\bibitem{Barbieri2} R. Barbieri, L. J. Hall and A. Strumia, {\it Nucl. Phys.}
{\bf B445} 219, (1995).
%
\bibitem{Barbieri3} R. Barbieri, L. J. Hall and A. Strumia,
{\it Nucl. Phys.} {\bf B449} 437, (1995).
%
\bibitem{Barbieri4} P. Ciafaloni, A. Romanino and A. Strumia,
IFUP-YH-42-95.
%
\bibitem{Barbieri5} T. V. Duong, B. Dutta and E. Keith, {\it Phys. Lett.}
{\bf B378} 128, (1996).
%
\bibitem{Barbieri6} G. Couture, et. al., {\it Eur. Phys. J.} {\bf C7}
135, (1999).
%
\bibitem{Barbieri7} Y. Okada, K. Okumara and Y. Shimizu, {\it Phys. Rev.}
{\bf D61} 094001, (2000).
%
\bibitem{Khalil} S. Khalil, {\it Phys. Rev.}
{\bf D81} 035002, (2010).
%
\bibitem{Chang} D. Chang, W. S. Hou and W. Y. Keung,
{\it Phys. Rev.} {\bf D48} 217, (1993).
%
\bibitem{Paradisi} P. Paradisi, {\it JHEP} {\bf 0602} 050, (2006).
%
\bibitem{MuLinYan} G. J. Ding, M. L. Yan,
{\it Phys. Rev.} {\bf D77} 014005, (2008).
%
\bibitem{AndiHektor} A. Hektor, Y. Kajiyama, K. Kannike,
{\it Phys. Rev.} {\bf D78} 053008, (2008).
%
\bibitem{Illana} F. del Aguila, J. I. Illana, M. D. Jenkins,
{\it JHEP} {\bf 0901} 080, (2009).
%
\bibitem{Illana2} J. I. Illana, M. D. Jenkins, {\it Acta Phys. Polon.}
{\bf B40} 3143, (2009).
%
\bibitem{Goto} T. Goto, Y. Okada,
Y. Yamamoto, {\it Phys. Rev.} {\bf D83} 053011, (2011).
%
\bibitem{Aguila} F. del Aguila, J. I. Illana, M. D. Jenkins,
{\it JHEP} {\bf 1103} 080, (2011).
%
\bibitem{GangHe} X. G. He, S. Oh, {\it JHEP} {\bf 0909} 027,
(2009).
%
\bibitem{Morisi} D. Ibanez, S. Morisi, J.W.F. Valle, {\it Phys. Rev.}
{\bf D80} 053015, (2009).
%
\bibitem{Hagedorn} C. Hagedorn, E. Molinaro, S. T. Petcov, {\it
JHEP} {\bf 1002} 047, (2010).
%
\bibitem{Deppisch} F. F. Deppisch, F. Plentinger, G. Seidl, {\it
JHEP} {\bf 1101} 004, (2011).
%
\bibitem{Ding} G. J. Ding, J. F. Liu ,  {\it JHEP} {\bf 1105} 029, (2010).
%
\bibitem{Aranda} J.I. Aranda, A. F. Tlalpa, F. R. Zavaleta , F. J. Tlachino,
J. J. Toscano, E. S. Tututi, {\it Phys. Rev.} {\bf D79} 093009,
(2009).
%
\bibitem{Akeroyd} A. G. Akeroyd, M. Aoki, H. Sugiyama, {\it Phys. Rev.}
{\bf D79} 113010, (2009).
%
\bibitem{Goudelis} A. Goudelis, O. Lebedev, J. H. Park, hep-ph/1111.1715,
(2011).
%
\bibitem{Tae}T. Hur, P. Ko, X. H. Wu, {\it Phys. Rev.}
 {\bf D76}, 096008 (2007)
%
\bibitem{IltanTensUnp} E. Iltan, {\it J. Phys. } {\bf G38  }, 105001  (2011).
%
\bibitem{Georgi1}H. Georgi, {\it Phys. Rev.  Lett.}  {\bf 98}, 221601 (2007).
%
\bibitem{Georgi2} H. Georgi, {\it Phys. Lett.} {\bf B650}, 275 (2007).
%
\bibitem{BankZaks} T. Banks, A. Zaks, {\it Nucl. Phys.} {\bf B196},
189 (1982).
%
\bibitem{Rajaraman} M. Bander, J. L. Feng, A. Rajaraman, Y.
Shirman, {\it Phys. Rev.} {\bf D76}, 115002 (2007).
%
\bibitem{PJFox}  P. J. Fox, A. Rajaraman, Y. Shirman,
{\it Phys. Rev.}  {\bf D76}, 075004 (2007).
%
\bibitem{Kikuchi} T. Kikuchi, N. Okada, {\it Phys. Lett.} {\bf B661}, 360
(2008).
%
\bibitem{Grinstein}  B. Grinstein,  K. A. Intriligator, I. Z. Rothstein
{\it Phys. Lett.} {\bf B662}, 367 (2008).
%
\bibitem{Cheung2} K. Cheung, W. Y. Keung and T. C. Yuan,
{\it Phys. Rev.}  {\bf D76}, 055003 (2007)..
%
\bibitem{ARajaraman}  A. Rajaraman, {\it Phys. Lett.} {\bf B671}, 411
(2009).
%
\bibitem{ADelgado}A. Delgado, J. R. Espinosa, J. M. No and M. Quiros,
{\it JHEP} {\bf 0804}, 028 (2008)


%
%
\end{thebibliography}
\end{document}